\documentstyle[12pt]{article}
\begin{document}
\hsize = 6.0 in
\vsize =11.7 in
\hoffset=0.1 in
\voffset=-0.5 in
\baselineskip=20pt
\newcommand{\ghat}{{\hat{g}}}
\newcommand{\Rhat}{{\hat{R}}}
\newcommand{\ih}{{i\over\hbar}}
\newcommand{\Scal}{{\cal S}}
\newcommand{\fudge}{{1\over16\pi G}}
\newcommand{\tn}{\mbox{${\tilde n}$}}
\newcommand{\mg}{\mbox{${m_g}^2$}}
\newcommand{\mf}{\mbox{${m_f}^2$}}
\newcommand{\hk}{\mbox{${\hat K}$}}
\newcommand{\vk}{\mbox{${\vec k}^2$}}
\newcommand{\eqletter}{ \hfill (\theequation\alph{letter})}
\newcommand{\gm}{{(\Box+e^2\rho^2)}}
\newcommand{\eql}{\nonumber &\eqletter \\
                  \addtocounter{letter}{1}}
\newcommand{\be}{\begin{equation}}
\newcommand{\ee}{\end{equation}}
\newcommand{\bea}{\begin{eqnarray}}
\newcommand{\eea}{\end{eqnarray}}
\newcommand{\beal}{\setcounter{letter}{1} \begin{eqnarray}}
\newcommand{\eeal}{\addtocounter{equation}{1} \end{eqnarray}}
\newcommand{\none}{\nonumber \\}
\newcommand{\req}[1]{Eq.(\ref{#1})}
\newcommand{\reqs}[1]{Eqs.(\ref{#1})}
\newcommand{\larrow}{\,\,\,\,\hbox to 30pt{\rightarrowfill}
\,\,\,\,}
\newcommand{\slarrow}{\,\,\,\hbox to 20pt{\rightarrowfill}
\,\,\,}
\newcommand{\half}{{1\over2}}
\newcommand{\bfx}{{\vec{x}}}
\newcommand{\bfy}{{\vec{y}}}
\newcommand{\zfp}{Z_{{FP}}}
\newcommand{\zf}{Z_{{F}}}
\newcommand{\zr}{Z_{{R}}}
\newcommand{\zop}{Z_{{OP}}}
\newcommand{\zekt}{Z_{EKT}}
\newcommand{\phstar}{{\varphi^\dagger}}

\def\DA{D_{(A)}}
\def\S3{S^{(3)}}
\def\half{{1\over2}}
\def\algso3{{\cal L}_{SO(3)}}

\begin{center}
{\bf The Geometry and Topology of 3-Manifolds and Gravity}\\
\vspace{15 pt}
{\it by}\\
\vspace{13 pt}
J. Gegenberg \\[5pt]
{\it Department of Mathematics and Statistics}\\
   {\it University of New Brunswick}\\
   {\it Fredericton, New Brunswick}\\
   {\it CANADA E3B 5A3}\\
e-mail address:  jack@math.unb.ca \\[5pt]
\end{center}
\vspace{40pt}
{\narrower\smallskip\noindent
{\bf Abstract} :
It is well known that one can parameterize 2-D Riemannian
structures by conformal transformations and diffeomorphisms of
fiducial constant curvature geometries; and that this construction
has a natural setting in general relativity theory in 2-D.  I will
show that a similar parameterization exists for 3-D Riemannian
structures, with the conformal transformations and diffeomorphisms of
the 2-D case replaced by a finite dimensional group of gauge
transformations.  This parameterization emerges from the theory of
3-D gravity coupled to topological matter.
\bigskip
\begin{center}March 1993\end{center}
\bigskip \noindent UNB Technical Report 93-01
\clearpage
\section{Introduction}
In the the late 1970's Thurston\cite{thurston78} proposed a {\it
geometric} classification of the topologies of closed three
dimensional manifolds.  In his scheme, each such manifold decomposes
into a connected sum of simpler manifolds $M=M_1\sharp M_2\sharp\cdot
\cdot \cdot$, obtained by cutting along two-spheres and tori.  These
simpler manifolds are conjectured to admit one and only one of eight
possible {\it geometric structures}.  Although this geometrization
conjecture is to date unproved\footnote{However, an interesting
possible approach to its proof, proposed by R. Hamilton and by
Isenberg and Jackson\cite{flow}, uses Ricci flows and the technology
of Bianchi models in relativistic cosmology.}, it has led to
significant advances in our understanding of three dimensional
geometry and topology, especially hyperbolic geometry and topology.
For a review see \cite{thurston78}.

It is clear that Thurston's scheme should be relevant to fundamental issues
in theoretical physics.  In particular, if the quantization of
gravity does not "fix" spatial topology, then in the functional
integral approach to calculating amplitudes we should integrate over
spatial topologies.  Indeed, in quantum cosmology, there are
interesting calculations involving such integrals.  See the work of
Fujiwara's group \cite{fuji} and Carlip \cite{carlip1}.  Furthermore,
it has been known for some time that locally homogeneous cosmological
models -the Bianchi/Kantowski-Sachs models- are essentially based on
the eight Thurston geometries refered to above
\cite{fagundes}\cite{ash/sam}.

In the following, I will show first of all, that the Thurston
geometries are fiducial solutions of a three dimensional gravity
theory, in the same way that the constant curvature
geometries are fiducial solutions of two dimensional Einstein
gravity.  Secondly, it will be shown that this leads to an
alternative characterization of three dimensional manifolds in terms
of flat bundles with structure group SO(3) over the three manifold.
The upshot is that one may then parameterize all the
Riemannian metrics on the given 3-manifold in terms of gauge
transformations of the fiducial metric of the appropriate Thurston
geometry.

In the remainder of this introduction we will show how two
dimensional gravity theory provides a natural physical setting for
the geometric structure of two dimensional manifolds.  In order to
accomplish this, we will first review geometric structures in
general, then briefly review the classification of closed two
dimensional manifolds in terms of the three geometric structures
associated with the manifolds of positive, zero and negative constant
curvature.  In section 2 we will review Thurston's eight geometric
structures and their relation to the question of the classification
of three dimensional manifolds-the ``geometrization conjecture".  In
section 3, after a brief explication of the properties of a relevant
topological field theory of gravity interacting with topological
matter, we will argue that the latter is a natural setting for
Thurston's geometrization conjecture and propose a new construction
of the geometric structure of locally homogeneous three manifolds.

An $(X,G)$ {\it structure} on a manifold $M$ is a pair $(X,G)$ where
$X$ is a manifold with dim$X=$dim$M$ and $G$ is a Lie group acting
transitively on $X$, such that $M$ is covered with "charts"
$\bigl\{\left(\phi_\alpha, U_\alpha\right)\bigr\}$ such that on the
overlap $U_\alpha\cap U_\beta\not=\emptyset$, the map $\phi
_\alpha\circ\phi^{-1}_\beta\in G$.\cite{goldman1}

A {\it locally homogeneous structure} on $M$ is an $(X,Isom(X))$ structure
on $M$, where
$X$ is a Riemannian manifold and $Isom(X)$ is the group of isometries
admitted by $X$.

It has been known for a long time \cite{2d} that a given closed
orientable two dimensional manifold admits one and only one of the
following three locally homogeneous structures:

$$\left(S^2,
Isom(S^2)\right), \left(T^2,Isom(T^2)\right),
\left(H^2,Isom(H^2)\right).$$
The manifolds $S^2, T^2, H^2$ are,
respectively, the closed orientable two dimensional manifolds of
constant positive, zero and negative curvature-i.e.  the sphere, the
(flat) torus and the (closed) hyperbolic space.  The groups
$Isom(S^2)$, etc., are the three dimensional groups of isometries of
the sphere, etc.  In the sense of F. Klein, these locally homogeneous
structures are called {\it geometries}.  For a given geometry, the
allowed topologies are determined by the set of finite subgroups of
the corresponding isometry group.

Furthermore, there is a fiducial constant curvature Riemannian metric
$\hat g$ for a given geometry and topology.  This fiducial metric can
be written in terms of the modular parameters associated with the
complex structure of the Riemann surface form of the manifold.
Locally, a Riemannian metric is determined by three smooth functions
of the coordinates.
  Hence a {\it given} Riemannian metric $g$ on the manifold can be
  written:
\be
g=e^{2\sigma}\Phi_*(\hat g),
\ee
where $\sigma$ is a
function on the manifold and $\Phi$ is a diffeomorphism.  We may
view, at least formally, the fiducial metric $\hat g$ as the
equivalence class of the Riemannian metrics $g$ {\it modulo} the
action of multiplication by a conformal factor $e^{2\sigma}$ and the
diffeomorphism group.

At first sight this seems wrong:  it would seem to lump into the same
equivalence class two distinct geometries, for example a flat metric
$g_{(0)}=\delta_{ij}dx^i\otimes dx^j$ and a metric of constant
positive curvature $g_{(+)}=(1- r^2/4)^{-2}g_{(0)}$, where
$r^2:=\delta_{ij}x^i x^j$.  It is true that in a given coordinate
patch $U$ these two metrics are conformally related, but this cannot
be extended to the whole manifold-the conformal factor
$(1-r^2/4)^{-2}$ is not defined at the points where $r^2=4$.

This emerges quite naturally from two dimensional general relativity
theory.  We write the action functional for that theory in first
order (``Palatini") form.  The fields are a dyad $e^a_\mu$ and an
SO(2) connection $\omega^a_{b\mu}$.  The action functional is:

\be
S^{(2)}={1\over2\pi}\int_{M^2}d^2x e \delta_{ab}e^a_\mu e^b_\nu
R^{\mu\nu}(\omega).
\ee
In the above, $e$ is the determinant of
$e^a_\mu$ and $R^{\mu\nu}(\omega)$ is the Ricci tensor constructed
from the connection $\omega^a_{b\nu}$.  The equations of motion,
obtained by requiring that $S^{(2)}$ is stationary under variations
of $e$ and $\omega$, fix the compatibility of the connection
$\omega^a_{b\mu}$ with an {\it arbitrary} smooth dyad $e^a_\nu$.  In
fact, $S^{(2)}$ is a topological invariant of the manifold $M^2$- it
is the Euler number $\chi(M^2)$.  This fact effectively fixes the way that
the locally arbitrary dyads "glue together" in overlapping patches so that
the geometry is ``compatible" with a fiducial geometry appropriate to the
topology of $M^2$.  By ``compatible" here, I mean that the metric is of the
form of Eqn. (1).

The topological invariance of $S^{(2)}$ can be viewed as
a consequence of the ``gauge invariance" of $S^{(2)}$ under conformal
transformations and diffeomorphisms.  By conformal transformations, I
mean transformations of the dyad field components of the form:
\be
e^a_\mu(x)\rightarrow e^{\sigma(x)}e^a_\mu(x).
\ee
Under such transformations, the integrand in the action is mapped to
itself plus a total derivative, and since $M^2$ is compact, this term
vanishes by the smoothness of the conformal factor $e^\sigma$.\footnote{One
might think that this argument could also apply to $S^{(2)}$ itself, since
the latter can be written in term of differential forms as $\int_{M^2}d
\omega$, where $\omega$ is the connection 1-form.  The point is that $\omega$
is not globally defined.  Alternatively, fiducial constant curvature metrics
$\hat g$ can be written as a "conformal factor" times a flat metric only on
a patch-the "conformal factor" is not globally defined.}

\bigskip
\section{Three-Manifold Geometries}

The situation for closed orientable 3-manifolds is complicated first
of all by the fact that there are locally homogeneous structures on
3-manifolds that are not isometric to one of the three constant
curvature spaces:  $S^3, E^3, H^3$.  In Table 1. below, the eight
3-manifold geometries are listed, specifying the underlying manifold,
the isometry group and a fiducial Riemannian metric admitting the
corresponding isometry group
\cite{thurston82}\cite{scott}\cite{fagundes}.
\clearpage
\begin{center}{\bf Table 1.}\end{center}
\vskip 5 pt
\[\begin{array}{rcl}
Manifold & Isometry Group &
Metric\\
 \;&\;&\; \\
S^3 & SO(4) & \cos^2ydx^2+dy^2+(dz-\sin ydx)^2 \\
E^3 & R^3\times SO(3) & dx^2+dy^2+dz^2 \\
H^3 & PSL(2,C) & dx^2+e^{2x}\left(dy^2+dz^2\right) \\
S^2\times E^1 & \left(Isom(S^2)\times Isom(E^1)\right)^+ & dx^2+dy^2+\sin^2y
dz^2 \\
H^2\times E^1 & \left(Isom(H^2)\times Isom(E^2)\right)^+ & dx^2+dy^2+e^{2x}dz^2
\\
\widetilde{SL(2,R)} & Isom(H^2)\times R & \cosh^2y dx^2+dy^2+(dz+\sinh y dx)^2
\\
Nil & Isom(E^2)\times R & dx^2+dy^2+(dz-xdy)^2\\
Sol & Sol\times (Z_2)^2 & dx^2+e^{-2x}dy^2 +e^{2x}dz^2
\end{array}\]
In the table, the seventh and eighth geometries are the most obscure.  The
3-manifold $Nil$ is a twisted product of $E^1$ with $E^2$.  Alternatively,
$Nil$ is the manifold of the Heisenberg group, i.e., the group of matrices:
\[\left\{\left(
\begin{array}{ccc}
1 & x & z \\
0 & 1 & y \\
0 & 0 & 1
\end{array}
\right)\right\}.\]
The 3-manifold $Sol$ is the solvable Lie group.  It is the only geometry
whose isometry group is three dimensional-its isotropy subgroup is trivial.
The metrics in Table 1. were obtained by Fagundes \cite{fagundes}, and are
the standard metrics for the three dimensional space sections of the Bianchi/
Kantowski-Sachs models associated with the given geometries.

The further complication is the fact, contrary to the 2-D case, that
direct sums of two or more of these geometries may not admit one of
the eight geometries.  Thurston has conjectured that any closed
orientable 3-manifold
 can be decomposed into components:

$$M=M_1\sharp M_2\sharp\cdot\cdot\cdot,$$
such that each of the $M_1,M_2,$ etc. admit one and only one of the
eight geometries.  This decomposition consists of cutting $M$ along
2-spheres and tori and gluing 3-balls to the resulting boundary
spheres on each piece.  This conjecture has not been proved to date.
However large classes of 3-manifolds have been shown to obey
it\cite{scott}.


\bigskip
\section{Three Dimensional Gravity}

In Chapter 1. we saw that two dimensional Einstein gravity theory, while
essentially a trivial theory physically, nevertheless encodes the structure
of two dimensional geometry and topology.  It would be nice if this were to
be the case in three dimensions, but it is not.  The reason for this is as
follows.  The Einstein-Hilbert action functional in three dimensions is:
\be
S_0^{(3)}=\int_{M^3} d^3x \sqrt{g} g^{\mu\nu} R_{\mu\nu}.
\ee
Its stationary points are the {\it flat} Riemannian geometries on
$M^3$.  One of the eight 3-manifold geometries, namely
$\left(E^3, R^3\times SO(3)\right)$, is a stationary point.  The other two
maximally symmetric geometries, namely $\left(S^3, SO(4)\right)$ and
$\left(H^3, PSL(2,C)\right)$ are the stationary points of action functionals
obtained from $S^{(3)}_0$ above by adding ``cosmological constant terms" of the
form
$\Lambda\sqrt{g}$ to the integrand.  There does not appear to be any simple
prescriptions for constructing action functionals whose stationary points are,
respectively, the remaining five {\it anisotropic} geometries.   This is
in contrast to the two dimensional case where essentially all Riemannian
geometries are stationary points of some simple action functional.

The action functional for three dimensional gravity originates from a
topological
invariant- the Chern-Simons invariant-
as was shown by
Achucarro and Townsend\cite{ach} and rediscovered later by E. Witten\cite
{witten}.   However, the Chern-Simons invariant does not encode the same
topological information as does the Einstein-Hilbert action functional in
two dimensions.  In fact the latter, as was mentioned above, is the Euler
number of the manifold; in the case of closed compact three dimensional
manifolds, the Euler number is zero.\footnote{In two dimensions, there is
a functional analogous to the Chern-Simons form in three dimensions\cite
{jackiw}\cite{teitelboim}\cite{cham}\cite{isler}.  It is the action
functional for a two-dimensional topological field theory of the so-called
BF type.}

I will argue here that the three dimensional theory of gravity
interacting with topological matter, developed in collaboration with
S. Carlip\cite{cargeg} in 1991, encodes the geometry and topology of
3-manifolds in manner strongly analogous to the situation of Einstein
gravity and 2-manifolds.  We suppose that $M^3$ is a smooth
orientable closed compact 3-manifold.  The cotangent bundle to $M^3$,
$T^*M^3$, is a fiber bundle over $M^3$ with structure group SO(3).  The
fibers are three dimensional vector spaces which come equipped with a
``natural" metric $\delta_{ab}$ and volume element $\epsilon_{abc}$.
A smooth frame field on $M^3$ is a set of three independent 1-form
fields $E^a$ on $M^3$.  A spin connection $A_a$ on $M^3$ is an
SO(3) connection on $M^3$.  The cotangent bundle $T^*M^3$ has
fibers isomorphic to the Lie algebra $\algso3$ of SO(3).  A spin
connection $A_a$ is compatible with a frame field $E^a$ if:
\be
D_{(A)}E^a:=dE^a+\half\epsilon^{ab}_cA_b E^c=0.
\ee
In the above, $D_{(A)}$ is the covariant exterior derivative with respect to
the connection $A_a$.  The $\algso3$ indices $a,b,...$ are raised and lowered
by the metric $\delta_{ab}$.  A spin connection and compatible frame field
determine
a Riemannian metric $g:=\delta_{ab}E^a\otimes E^b$ on $TM^3$.

Let $E^a, B^a, C_a$ be three sets of 1- form fields over $M^3$.  The fields
need not be non-degenerate nor mutually linearly independent, nor
compatible with connection $A^a$.
The action is a
functional of the spin connection $A_a$ and the 1-form fields
$B^a,C_a, E^a$:
\be
\S3= \half\int_{M^3}E^a\wedge F_a(A)+B^a\wedge \DA
C_a.
\ee
The curvature $F_a(A)$ of the connection $A_a$ is
$$F_a(A):=dA_a+{1\over4}\epsilon^{bc}_a A_b A_c.$$

The stationary points of $\S3$ are given by the solutions of the following
set of first order partial differential equations:
\bea F_a(A)&=0,\label{eq:7a} \\
\DA B^a&=0,\label{eq:7b} \\
\DA C_a&=0,\label{eq:7c} \\
\DA E^a+\half \epsilon^{abc}B_b\wedge C_c&=0.\label{eq:7d}
\eea
In general, the three 1-form fields $E^a$ are not a frame field compatible with
the spin connection $A_a$.  This is because of the term in $B_b\wedge C_c$
in the last equation of motion.

Nevertheless, the equations of motion above determine a Riemannian geometry
on $TM^3$ as follows.  Consider the 1-form field $Q_a$ satisfying
\be
\epsilon^{abc}\left(Q_b\wedge E_c -B_b\wedge C_c\right)=0.
\ee
Then the equation of motion for the $E^a$ can be written as:
\be
dE^a+\half\epsilon^{abc}\left(A_b+Q_b\right)\wedge E_c=0.
\ee
This is of the form of the condition that the frame field $E^a$ is compatible
with the connection $\omega_a$ defined by:
\be
\omega_a:=A_a+Q_a.
\ee

The following Theorem is straightforward to prove:
\par\noindent
{\it Theorem}:  Each of the eight geometries of Table 1. is a stationary point
of the action functional $\S3$.
\par\noindent
{\it Proof}:  We prove the theorem by constructing solutions of
Eqns.(7) on a
coordinate
patch $U$ of $M^3$.  Since $A_a$ is flat, we can choose a gauge so
that $A_a=0$ on $U$.  In Table 2. below I display for each of the eight
3-manifold geometries the corresponding frame field $E^a$ (obtained by
simply factoring the fiducial metric given in Table 1.) and closed 1-form
fields $B^a, C_a$ such that that the last of the equations of motion in
Eqns(7) is satisfied.  The compatible spin connection for each geometry is
precicely that given by Eqns(8)-(10).
\clearpage

The action functional $\S3$ is invariant under the group whose infinitesitmal
generators are given by the following \cite{cargeg}:
\bea\delta B^a=&\DA \rho^a+\half\epsilon^{abc}B_b\tau_c,\label{eq:11a}\\
\delta C^a=&\DA\lambda^a+\half\epsilon^{abc}C_b\tau_c,\label{eq:11b}\\
\delta E^a=&\DA\xi^a+\half\epsilon^{abc}\left(E_b\tau_c+B_b\lambda_c+C_b
\rho_c\right),\label{eq:11c}\\
\delta A^a=&\DA\tau^a,\label{eq:11d} \eea
where the twelve quantities $\tau^a,\lambda^a,\rho^a,\xi^a$ are infinitesimal
parameters generating the transformations.    The group generated by these
infinitesimal transformations is I(ISO(3)).  The notation IG denotes the
group obtained by taking the semi-direct product of the Lie Group G with its
own Lie algebra ${\cal L}_G$.

I now conjecture that given the topology of a prime manifold $M^3$,
most Riemannian metrics on $M^3$ can be parameterized by the gauge
parameters of a gauge
transformation of the form of Eqns.(16).
We first note the following \cite{cargeg}:  (i.)  The action
Eq. (6) is invariant up to a total divergence under gauge
transformations with {\it finite} values of the gauge parameters
$\xi^a,\rho^a,\lambda^a$ as long as $\tau^a=0$;  (ii.)  For any
topology, the equations of motion admit the trivial configuration
$A^a=B^a=C^a=E^a=0$.
\par
We can obtain the following configuration, which is a solution of
the equations of motion, by performing the following finite gauge
transformations in succession on the trivial configuration
\cite{cargeg}:
\bea
\delta_1\left(B^a,C^a,E^a\right)&=&\left(d\rho^a,0,d\xi^a\right), \nonumber \\
\delta_2\left(B^a,C^a,E^a\right)&=&\left(0,d\lambda^a,\half\epsilon^{abc}\delta_1 B_b\lambda_c\right),
\label{eq:sucgauge}
\eea
to get
\be
E^a=d\xi^a+\half\epsilon^{abc}d\rho_b\lambda_c,
\label{eq:Egauge}
\ee
\bea
B^a&=d\rho^a, \\
C^a&=d\lambda^a.
\eea
\par
By the appropriate choice of gauge parameters, we can get the
fiducial $E^a$ in Table 2.  In a sense to be enlarged on below, the
triad $E^a$ defined by \req{eq:Egauge} is general.  Consider the
choice of parameters $\xi^A=0, \rho^1=0, \rho^A=x^A$ with $A=2,3$.
Then the metric $g_{\mu\nu}=\delta_{ab}E^a_\mu E^b_\nu$ can be
written in "ADM" form:
\be
ds^2=N^2(dx^1)^2+h_{AB}\left(dx^A+N^A
dx^1\right)\left(dx^B+N^Bdx^1\right),\label{eq:adm}
\ee
where
$$N^2={\tilde\lambda_3\over 2}\partial_1\xi^1, \;\; N^3=-
{\tilde\lambda_2\over 2}\partial_1\xi_1,\;\;
N={\lambda_1\over\tilde\lambda}\partial_1\xi^1,
$$
with
\begin{eqnarray*}
\tilde\lambda_A&:=\lambda_A+2\epsilon_{AB}\partial_B\xi^1, \\
\tilde\lambda&:=\sqrt{\lambda^2_1-\tilde\lambda^2_2-
\tilde\lambda^2_3},
\end{eqnarray*}
and
\be
h_{AB}dx^Adx^B=e^{2\phi}\big\vert d\omega+\mu
d\overline\omega\big\vert^2,
\nonumber \ee
with
\bea
e^\phi&:={1\over4}(\lambda_1+\tilde\lambda),\nonumber \\
\mu&:={e^{-2\phi}\over 16}(\tilde\lambda_2+i\tilde\lambda_3)^2
\label{eq:beltrami}
\eea
\par
The $E^3, H^3, S^2\times E^1, H^2\times E^1$ and $Sol$ geometries
can be so paramterized.  The others likely can, though not with the
particular metric triad given in Table 1. and 2.  However, at least
$S^3$ with metric/triad as in these tables can be parameterized by
\req{eq:Egauge} but now with the gauge parameters chosen as:
\begin{eqnarray*}
(\rho^a)&=(x,\ln\vert\csc{y}-\cot{y}\vert, x\cot{y}),\\
(\lambda^a)&=(0,-2\sin{y},2x), \\
(\xi^a)&=(0,y+{x^2\over 2}, z).
\end{eqnarray*}
\par
In the case of $S^2\times E^1$, with $x^2,x^3$, i.e.,
$\omega,\overline\omega$, coordinates on $S^2$-as argued in
\cite{cargeg}- {\it all} metrics on $S^2\times E^1$ are gauge
equivalent to the one which is gauged from the trivial
configuration.  Hence all metrics on $S^2\times E^1$ are gauge
equivalent to the fiducial metric given in Table 1.  This is
because all ``Beltrami differentials'' $\mu$ on $S^2$ are equivalent
up to diffeomorphisms \cite{baulieu}.
\par
For the other topologies the situation is rather subtle.  Consider
the case of $E^3$-i.e., $T^3$ since we are considering closed
compact topologies.  In this case $\mu$ fixes a point in
Teichm\"uller space, up to diffeomorphisms.  Hence a given geometry
$S^1 \times T^2_{[\mu]}$, with the subscript $[\mu]$ denoting the
Teichm\"uller parameters of the torus, is gauge equivalent to the
trivial configuration with the gauge parameters constrained by
\req{eq:beltrami}.  All such geometries are then gauge equivalent
to the fiducial $T^3$ geometry with appropriate Teichm\"uller
parameters.  A more comprehensive analysis is currently underway.

Furthermore, I
conjecture that each of the 3-manifold geometries is characterized by
a flat SO(3) connection $A_a$ {\it modulo} gauge equivalence under
the group I(ISO(3)), and two $\algso3$-valued 1-form fields $B^a,C_a$
closed with respect to the flat connection $A_a$.

Finally, it is worth noting here
that this structure somewhat resembles Thurston's characterization of
geometric structures in terms of a flat bundle equipped with transverse
foliation and canonical section \cite{thurston83}\cite{goldman1}.

The proof of these conjectures is currently under investigation.
\par\noindent
{\bf Acknowledgments} \\
I wish to thank Yongyi Bi, Steve Carlip and Robert Meyerhoff for
useful conversations.  I also wish to acknowledge the partial support
of the Natural Sciences and Engineering Research Council of Canada for
this research.

\end{document}